\documentclass[prb,twocolumn,floatfix,showpacs,superscriptaddress]{revtex4}
\usepackage{bm}
\usepackage{amsmath}
\usepackage{amssymb}
\usepackage[dvips]{color}
\usepackage[dvips]{graphics}
\usepackage[dvips]{graphicx}
\usepackage{eepic}
\usepackage{pstricks}%
\usepackage{pst-eps}%
\usepackage{pst-plot}%
\usepackage{pst-node}%
\usepackage{pst-poly}

\renewcommand{\i}{{\rm i}}
\newcommand{\e}{{\rm e}}

\begin{document}

\title{Link between the hierarchy of fractional quantum Hall states and Haldane's
conjecture for quantum spin chains}
\author{Masaaki Nakamura}
\affiliation{
Department of Physics, Tokyo Institute of Technology,
Tokyo 152-8551, Japan}
\author{Emil J. Bergholtz}
\affiliation{
Max-Planck-Institut f\"{u}r Physik komplexer Systeme,
N\"{o}thnitzer Stra{\ss}e 38, D-01187 Dresden, Germany}
\author{Juha Suorsa}
\affiliation{
Department of Physics, University of Oslo, P.O. Box 1048 Blindern, 0316
Oslo, Norway}

\date{\today}
\begin{abstract}
We study a strong coupling expansion of the $\nu=1/3$ fractional quantum
Hall state away from the Tao-Thouless limit and show that the leading
quantum fluctuations lead to an effective spin-$1$ Hamiltonian that
lacks parity symmetry.  By analyzing the energetics, discrete symmetries
of low-lying excitations, and string order parameters, we demonstrate
that the $\nu=1/3$ fractional quantum Hall state is adiabatically
connected to both Haldane and large-$D$ phases, and is characterized by
a string order parameter which is dual to the ordinary one.  This result
indicates a close relation between (a generalized form of) the Haldane
conjecture for spin chains and the fractional quantum Hall effect.
\end{abstract}

\maketitle

\section{Introduction}

There are striking similarities between the catalogue of $SU(2)$-symmetric
quantum spin chains\cite{Haldane1983} and the hierarchy of fractional
quantum Hall (FQH) states \cite{Laughlin,haldane83,halperin84,jain89}.
Arguably the most striking parallel is that
both systems allow a $\mathbb{Z}_2$ classification.
Haldane conjectured \cite{Haldane1983} that
half-integer $SU(2)$ quantum spin chains support
gapless excitations, protected by a topological term in the effective
action, while the integer spin chains develop a mass gap.  
A similar structure appears in the quantized FQH effect.
At filling factors $\nu<1$, quantized conductance plateaus 
only occur at rational $\nu$ with \emph{odd} denominator, 
while in the vicinity of \emph{even}-denominator fractions metallic behavior
is sustained. 
The Haldane conjecture and the phenomenology of the FQH effect
communicate something pivotal about the low-lying excitations of 
seemingly disparate quantum phases, of low-dimensional magnetic materials
and 2D electron gas in magnetic field. Hence, it is important
to establish whether the similarities
are merely accidental or if 
the structure of low-energy excitations in these systems have a
related microscopic origin.

\begin{figure}[t]
\includegraphics[width=7.5cm]{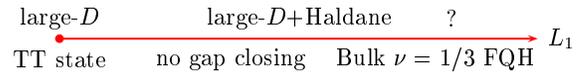}
\caption{(Color online) Phase diagram of $\nu=1/3$ FQH system as a
function of the circumference of the torus, $L_1$. The TT limit,
$L_1\rightarrow 0$, of the FQH problem corresponds to the large-$D$
phase of a spin-chain. By increasing $L_1$, the bulk FQH state is
adiabatically approached and the corresponding spin chain is, for
intermediate $L_1$, characterized by co-existing features of the
large-$D$ and Haldane phases. For very large $L_1$ the FQH/spin-chain
correspondence cannot be derived microscopically as indicated by
question mark on the spin-chain side of the phase
diagram.}\label{fig:pdiagram}
\end{figure}

Already two decades ago, a more precise analogy between the two systems
was discussed \cite{Girvin-A} in terms of off-diagonal long-range order
in FQH states \cite{Girvin-M} and hidden orders present in $S=1$ spin
chains \cite{Nijs-R} (see also
Refs. \onlinecite{Arovas-H,Nussinov-O,Arovas-Hasabe} for related
analogies). 
More recently, a natural framework for studying this connection
emerged as it was realized that universal features of
many quantum Hall (QH) phases are retained on a thin torus
\cite{Bergholtz-K2005,Seidel-F-L-L-M,Bergholtz-K2006-8,Bergholtz-H-H-K}
(or Tao-Thouless, TT \cite{Tao-T,Anderson,Bergholtz-K2006-8}) limit, 
where the interacting problem is trivially solvable.
FQH states at odd-denominator filling factor fraction can be deformed
into the TT limit without closing the energy gap, as has been
rigorously shown at the Laughlin fractions
\cite{Anderson,Rezayi-H,Bergholtz-K2005,Seidel-F-L-L-M,Bergholtz-K2006-8,Jansen-L-S}
and plausibly argued for at other fractions.\cite{Bergholtz-K2006-8,Bergholtz-H-H-K} 
Notably different
behavior is found in states at even-denominator filling.  For example,
analysis of gapless QH state at filling fraction $\nu=1/2$ shows that
the system undergoes a first order phase transition from a gapped TT
state to a gapless phase upon deformation of the
torus.\cite{Bergholtz-K2005,Bergholtz-K2006-8} In fact, this analysis
of the $\nu=1/2$ FQH state uses a $S=1/2$ spin chain description.  A
similar spin-chain picture\cite{Wikberg-B-K} can also capture features
of non-Abelian states\cite{mr}, which further adds to the analogies
between spin-chain physics and the FQH effect. The possibility of a
relationship between the Haldane conjecture and the FQH effect was
suggested in Ref.~\onlinecite{Bergholtz-K2005}. 
In this article, we
provide explicit evidence for such a link by obtaining the phase
diagram, sketched in Fig.~\ref{fig:pdiagram}, for the $\nu=1/3$ FQH
state away from the TT limit.
We also outline how such connection can be
extended to arbitrary filling fractions $\nu=p/q$. 

Our effective model for the $\nu=1/3$ system close to the TT limit
has the form of a parity breaking spin-$1$ chain. 
Recently, such models have
attracted interest in the context of ultra-cold lattice
bosons\cite{Berg-T-G-A} and are presently featuring in attempts to
generally classify topological phases\cite{Gu-W,Pollmann}. A consequence
of the lack of parity symmetry is that phases that are normally
separated by a phase boundary can co-exist. We demonstrate that the
effective spin model has characteristics of both the large-$D$ phase (in
which an anisotropic single-site term in the Hamiltonian, $H\sim
D\sum_i(S^z_i)^2$, freezes the spins into $S^z=0$) and the topologically
non-trivial Haldane phase\cite{Haldane1983}, and that the ground state
thereof is adiabatically connected (no gap closing in the thermodynamic
limit) to the ground states of both these phases. We also investigate to
what extent the character of the FQH state can be captured by string
order parameters, and find that a "dual" version of the conventional
string order parameter may be suitable in this context. The fact that
the gapped large-$D$ and Haldane phases both exclusively exist for
integer spin chains suggest that both play a role in a general
connection between quantum spin chains and hierarchical (Abelian) FQH
states.

The rest of this article is organized as follows. In Section \ref{sec2}
we study the FQH system away from the TT limit and motivate a parity
breaking spin-$1$ chain as an effective model of the $\nu=1/3$ system
and outline how this can be generalized to arbitrary fractions. In
Section \ref{sec3} we focus on the spin-$1$ ($\nu=1/3$) case and extend
the effective spin-model to enable interpolation to more conventional
spin models, which for example have large-$D$ and Haldane
phases. Concluding remarks are given in Section \ref{sec4}.

\section{Thin torus limit of the quantum Hall system}\label{sec2}

\subsection{Mapping to one-dimensional model}
We consider a model of $N$ interacting electrons in the lowest Landau
level on the torus. In the Landau gauge, a complete basis of $N_\phi$
degenerate single-particle states, labeled by $k=0,\ldots, N_\phi-1$,
can be chosen as
\begin{equation}
\psi_k(x)=
(\pi^{1/2} L_1)^{-\tfrac{1}{2}}\sum_{n=-\infty}^{\infty}
\e^{\i (k_1+n L_2) x_1}
\e^{-\tfrac{1}{2}(x_2+k_1+n L_2)^2},
\end{equation}
where $L_i$ are the circumferences of the torus, $x_i$ the corresponding
coordinates, and $k_1=2\pi k/L_1$ the momentum along the $L_1$-cycle.
We have set the magnetic length $l_{\rm B}\equiv \sqrt{\hbar/eB}$ equal
to unity.  In this basis, any translation-invariant 2D two-body
interaction Hamiltonian assumes the form
\begin{equation}
\mathcal{H}=\sum_{k>|m|}\hat V_{km},\quad
\hat V_{km}\equiv V_{km}
\sum_{i}
 c_{i+m}^{\dag}
 c_{i+k}^{\dag}
 c_{i+m+k}^{\mathstrut}
 c_{i}^{\mathstrut},
 \label{TT_model}
\end{equation}
where the matrix-element $V_{km}$ specifies the amplitude for a process
where particles with separation $k+m$ hop $m$ steps to a separation
$k-m$ (note that $m$ can be negative). At the filling $\nu=p/q$ the
Hamiltonian commutes with the center-of-mass magnetic
translations\cite{haldane85} $T_1$ and $T_2^q$ along the cycles, which
implies, in particular, that the total momentum $K$ along the
$L_1$-cycle is conserved modulo $N_\phi$ in this gauge.

Laughlin's state is an exact zero-energy eigenstate of the above
Hamiltonian with the choice
\begin{equation}
V_{km}=(k^2-m^2)\e^{-2\left(k^2+m^2\right)\pi^2/L_1^2}
\end{equation}
obtained as the matrix elements of a periodized Haldane pseudo-potential
$\nabla^2\delta(\bm{r}-\bm{r}')$.\cite{haldane83,Trugman-K} The
amplitudes $V_{km}$ are exponentially damped in $1/L_1^2$. Therefore, at
small $L_1$ the model can be approximated by a few most dominant terms
such as $\hat V_{10}$, $\hat V_{20}$, $\hat V_{21}$, etc. (From now on
refer to $\hat V_{km}+\hat V_{k,-m}$ as simply $\hat V_{km}$ for
brevity.) We also study the model with Coulomb matrix elements where
longer range electrostatic terms $\hat V_{k0}$ are non-negligible.

\subsection{Effective spin-$1$ model for $\nu=1/3$}
At the filling $\nu=N/N_\phi=1/3$, the ground state manifold of the
$\hat V_{10}$ and $\hat V_{20}$ -terms is three-fold degenerate, spanned
by charge ordered states with one electron per a three-site unit cell:
$|\cdots\ 010\ 010\ 010\ \cdots\rangle$.  The $\hat V_{21}$-term induces
fluctuations upon these ground states through the process
\begin{equation}
|010\ 010\rangle\leftrightarrow|001\ 100\rangle.
\end{equation}
The truncated model can be mapped to an $S=1$ quantum spin chain by
identifying the states of the unit cell as $|010\rangle\to|0\rangle$,
$|001\rangle\to|+\rangle$, and $|100\rangle\to|-\rangle$.  Clearly the
identification explicitly break translational symmetry---there are three
equivalent ways of grouping three electronic sites into one spin
site. By choosing a particular grouping of the sites (so that the ground
state appears at total $S^z=0$) we effectively mod out the original
three-fold degeneracy.\cite{worry} In terms of $S=1$ variables, the
$\hat V_{21}$-process is then accounted for by the Hamiltonian ${\cal
H}=\sum_{i=1}^N h_{i,i+1}$ with
\begin{align}
h_{ij}=&\frac{1}{2}S_i^+S_j^-
\left(1-(S_i^z)^2\right)\left(1-(S_j^z)^2\right)
+\mbox{H.c.}
\label{model.0}
\end{align}
We note that this
Hamiltonian does not have the space inversion and spin
reversal symmetries: the process $|00\rangle\leftrightarrow|+-\rangle$
exists but $|00\rangle\leftrightarrow|-+\rangle$ does not.
This ``parity'' breaking is a consequence of the
dependence of $V_{km}$ on the single-particle momentum transfer $m$. For
a fixed (initial) separation $k+m$, the amplitudes are asymmetric with
respect to $m\leftrightarrow -m$. Inward hops have a greater amplitude
than outward hops.

In the TT limit, the fractionalized excitations of the system are domain
walls between the degenerate vacua. These can be included in the
effective spin-chain description by introducing at the domain walls edge
spins that carry a lower, $S=1/2$, representation. The energetics of
spatially separated domain walls is not essential to the FQH
phenomenology as long as we can assume them to localize.  Hence, in this
paper we only analyze the exciton (bound quasielectron-quasihole pair)
gap, which we relate to the Haldane gap in the effective spin model.

\begin{figure}[h]
\begin{center}
\includegraphics[width=7.0cm]{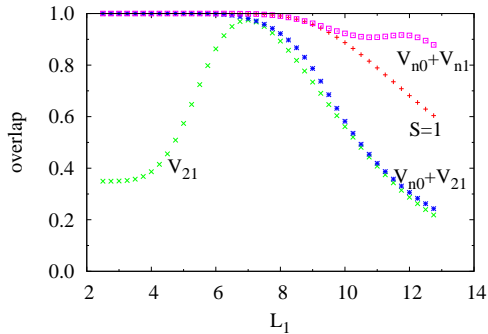}
\end{center}
\caption{(Color online) Overlaps of the exact $N=8$ Laughlin state with
the ground states of truncated Hamiltonians, consisting of a few leading
terms, as functions of $L_1$. Also shown is the projection of the exact
Laughlin state onto $S=1$-chain Hilbert space.}\label{fig:ovelaps}
\end{figure}

To study the relevance of the model (\ref{model.0}) for the Laughlin
state, we have analyzed ground state overlaps and excitation spectra.
Fig.~\ref{fig:ovelaps} shows as functions of $L_1$ the overlaps of the exact
Laughlin state, obtained as the ground state of the Hamiltonian
including all $\hat{V}_{km}$ terms, with the ground states of various truncations:
$\hat{V}_{21}$, $\sum_n \hat{V}_{n0}+\hat{V}_{21}$, $\sum_n
(\hat{V}_{n0}+\hat{V}_{n1})$. Also, the projection
of the Laughlin state onto the Hilbert space of $S=1$ chain is shown.

The high overlap with the $\hat{V}_{21}$ Hamiltonian at around $L_1=7$
indicates that the ground state around this $L_1$ is related to the ground
state of the $\hat{V}_{21}$-Hamiltonian, which when restricted to
the $S=1$ -chain Hilbert space, maps to the parity-broken
$S=1$ model in (\ref{model.0}). Further evidence comes 
from the fact that the truncated Hamiltonian reproduces the low energy part of the entanglement spectrum of the Laughlin state\cite{Lauchli}.

\begin{figure}[h]
\begin{center}
\includegraphics[width=7.0cm]{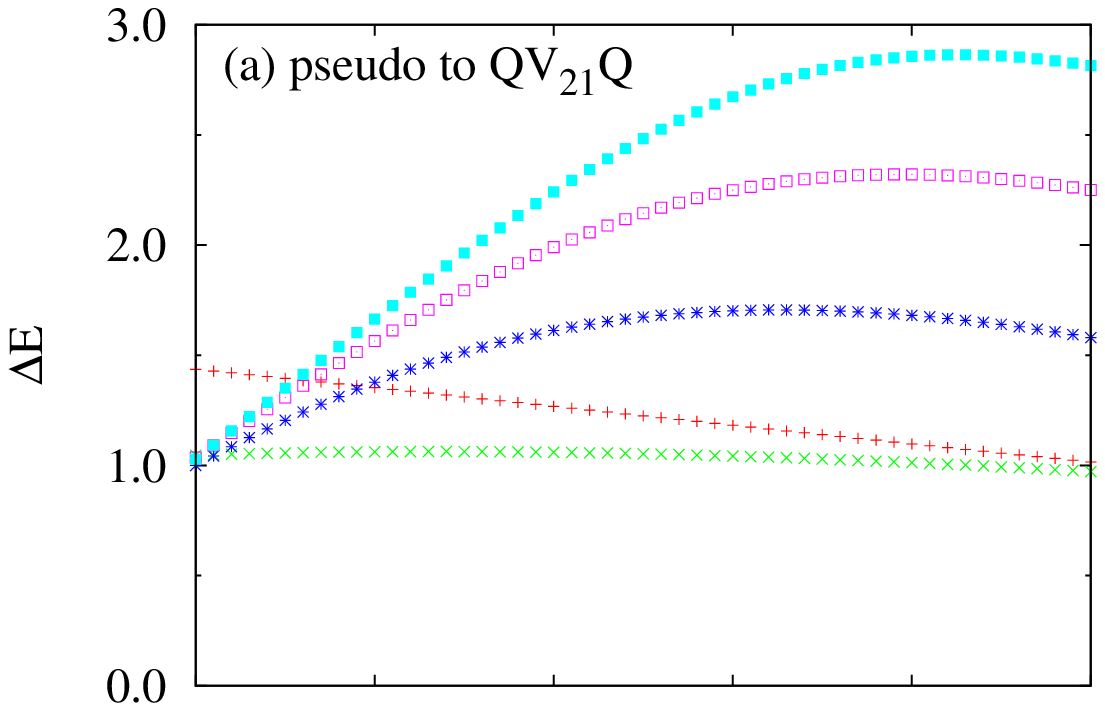}
\includegraphics[width=7.0cm]{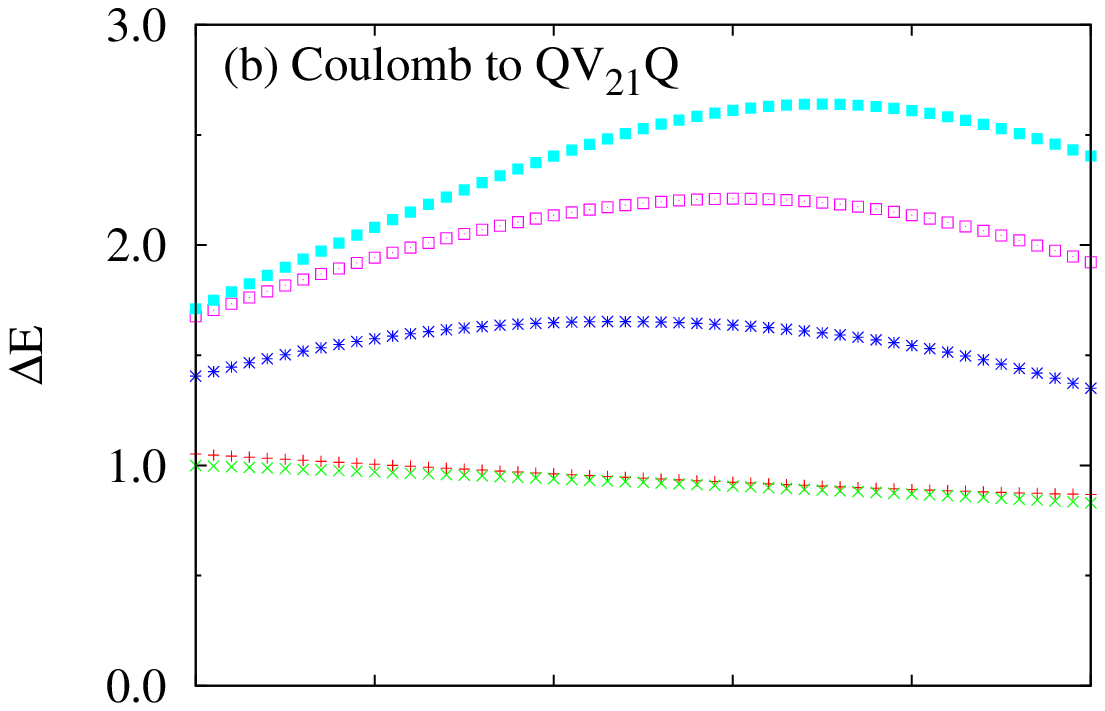}
\includegraphics[width=7.0cm]{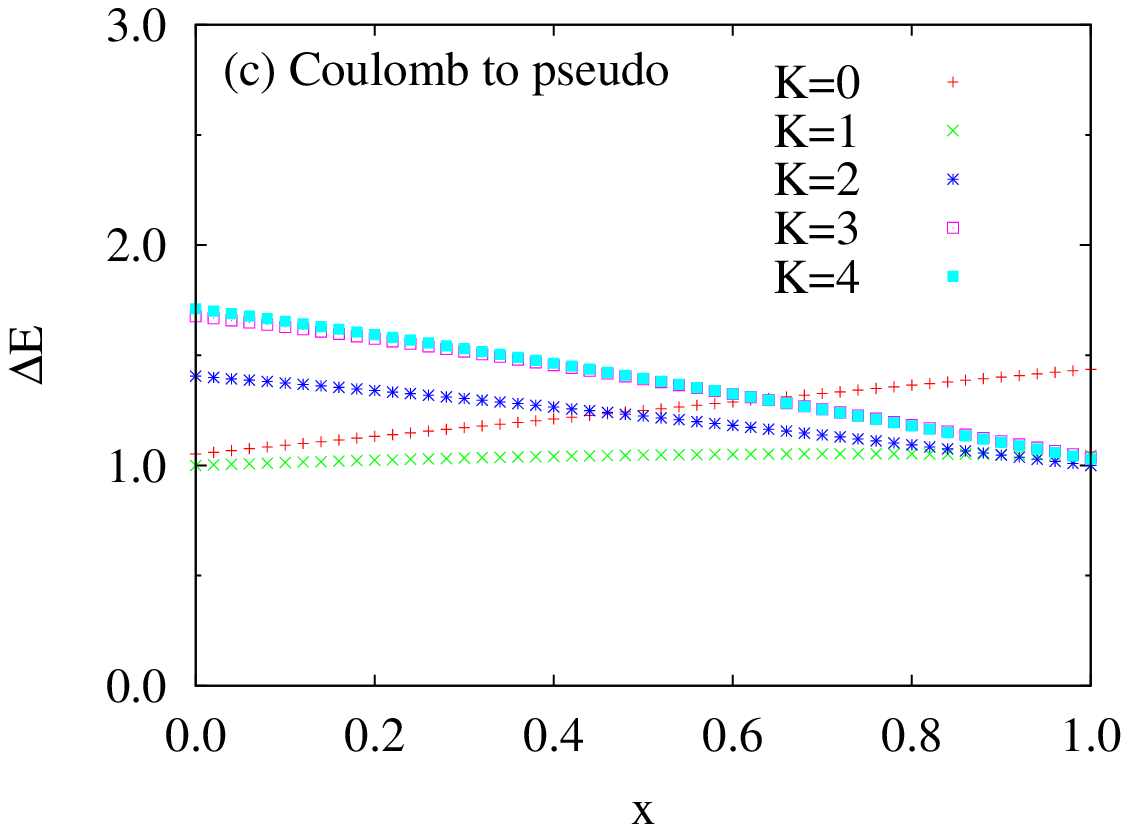}
\end{center}
\caption{(Color online) Excitation spectra (the lowest
levels for each $K$) of the Hamiltonians (a) ${\cal H}=(1-x){\cal
H}_{\textrm{p}}+x {\cal Q} \hat V_{21}{\cal Q}$, (b)${\cal H}=(1-x){\cal
H}_{\textrm{c}}+x {\cal Q}\hat V_{21}{\cal Q}$, and (c) ${\cal
H}=(1-x){\cal H}_{\textrm{c}}+x {\cal H}_{\textrm{p}}$, for fixed
$L_1=7$ and $N=8$.}\label{fig:gaps}
\end{figure}
To determine whether the low-lying excitations are also captured by a
$S=1$ spin chain, we study how the spectrum of the quantum Hall system
changes as we deform the potential from the full exact
pseudopotential and Coulomb potential to ${\cal Q}\hat{V}_{21}{\cal Q}$, where 
the projector ${\cal Q}$ projects to the Hilbert space of the $S=1$ chain.
Note that the pure hopping Hamiltonian $\hat{V}_{21}$ preserves $S=1$ Hilbert space
in the ground state sector, in which the TT state lies, but in general this is not 
true. For example, $\hat V_{21}$ acting on $100\, 100$ takes it to a 
configuration $011\, 000$, which lies outside the $S=1$ Hilbert space.
Hence, we include the projectors to make connection to the spin-chain models.
Results of the
analysis are shown in Fig.~\ref{fig:gaps}, where we have fixed
$L_1=7$. The panels show how the spectrum of lowest-lying
excitations in each $K$-sector changes upon various linear interpolations:
(a) full Coulomb Hamiltonian ${\cal H}_c$ to ${\cal Q}\hat{V}_{21}{\cal Q}$,
(b) ${\cal H}_c$ to the Haldane pseudopotential Hamiltonian
${\cal H}_p$,
and (c) ${\cal H}_p$ to ${\cal Q}\hat{V}_{21}{\cal Q}$. 
According to the discussion above our results provide an explicit
interpolation between the FQH Hamiltonian (\ref{TT_model}) and the spin
chain defined in (\ref{model.0}).\cite{projection} We observe that
the gap remains finite and approximately constant throughout the interpolations.
The ordering of the 
levels as a function of the momentum $K$ (relative to the ground state) 
can be thought of as the exciton dispersion. 
We find that the dispersions obtained with the Coulomb and the $
{\cal Q}\hat V_{21} {\cal Q}$ Hamiltonians largely agree.
It is interesting to note that the low-lying $K=0$ excitation 
crosses some of the finite-$K$ levels 
upon deformation of the Coulomb to the pseudopotential Hamiltonian. For the purpose of the present article we conclude that also the spectrum of our spin model is compatible with that of the FQH problem for a realistic interaction.

\begin{table}
\begin{tabular}{c|c|rrrrr}
& & ${\cal P}$ & ${\cal T}$ & $k$& BC & $M$\\ \hline
$E_0$ & G.S. & $+1$ &$+1$ & $0$ & $+1$ & $0$\\
$E_1$ & Haldane & $-1$& $-1$& $0$ & $-1$ & $0$\\
$E_2$ & Large-$D$ & $+1$ & $+1$ & $0$ & $-1$ & $0$\\
$E_3$ & Dimer & $+1$ &$+1$ & $\pi$ & $-1$ & $0$\\
$E_4$ & XY & $+1$ & $*$ & $0$ & $+1$ & $2$

\end{tabular}
\caption{Discrete symmetries of the excitation spectra (${\cal P}$:
space inversion, ${\cal T}$: spin reversal, $k$: wave number, and $M$:
total $S^z$). BC$=1$ (BC$=-1$) stands for (anti)periodic boundary
conditions. G.S. means the ground state.}  \label{tbl:symmetries}
\end{table}

\begin{figure}[b]
\begin{center}
\includegraphics[width=7.0cm]{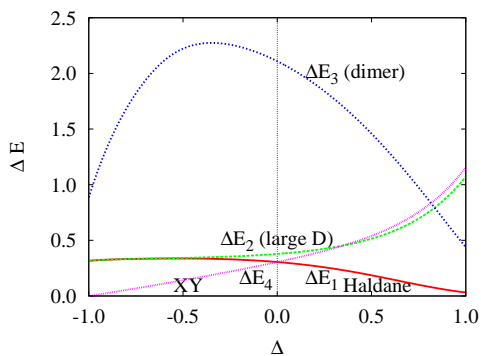} 
\end{center}
\caption{(Color online) Excitation spectra $\Delta E_i$ of the XXZ chain
($D=\lambda=0$) with system size $N=16$ under antiperiodic boundary
conditions. The lowest spectrum corresponds to different four ground
states (XY, Haldane, large-$D$, and dimer phases). A level crossing at
$\Delta=0$ corresponds to XY-Haldane phase transition point reflecting
the hidden SU(2) symmetry.\cite{Nomura-K,Kitazawa-H-N}}\label{dE_XXZ}
\end{figure}

\subsection{Effective spin model for generic filling fractions}

The mapping of the $\nu=1/3$ FQH system onto a spin-$1$ model carried
out above readily generalizes to arbitrary filling fractions. At
rational filling $\nu=p/q$ the TT ground states have unit cells of
length $q$ containing $p$ electrons being as far separated as possible
\cite{Bergholtz-K2006-8}. The $q$ degenerate translations of the unit
cell can be thought of as the $2S+1$ states of a spin $S=(q-1)/2$, which
suggests a mapping of the FQH system at the filling $\nu=p/q$ onto an
effective $S=(q-1)/2$ spin chain. This makes a general
connection between odd (even) denominator FQH fractions and the integer
(half-integer) spin-chains explicit.

\section{Analysis of the spin-$1$ model}
\label{sec3}

\subsection{Twisted boundary method for ground state}
In order to identify the universality class of the ground state of the
model (\ref{model.0}), we extend the Hamiltonian as
\begin{align}
 h_{ij}=& \frac{1}{2}
S_i^+S_j^-
 \left(1-\lambda(S_i^z)^2\right)\left(1-\lambda(S_j^z)^2\right)
 +\mbox{H.c.}\nonumber\\
&+\Delta S_i^zS_j^z
 +\frac{D}{2}\left((S_i^z)^2+(S_j^z)^2\right).
 \label{model}
\end{align}
We then study the adiabaticity of deformations from parameter regions
where physical properties are already known, to the point $\Delta=D=0$
and $\lambda=1$, which is related to the $\nu=1/3$ FQH effect according
to the discussion above, and will henceforth be referred to as the FQH
point.

Now let us review properties of this model (\ref{model}) for already
known parameter regions.  For $\lambda=D=0$, the model reduces to the
$S=1$ XXZ spin chain.  Then the system is ferromagnetic at
$\Delta<-1$. It has the XY phase at $-1\leq\Delta\leq0$, the Haldane
phase at $0<\Delta<\Delta_{\rm c}$, and is in the N\'{e}el state at
$\Delta_{\rm c}<\Delta$, where $\Delta_{\rm c}=1.17\pm
0.02$.\cite{Kitazawa-N} The XY-Haldane transition is of the
Berezinskii-Kosterlitz-Thouless (BKT) type, reflecting the SU(2)
symmetry of the XY model.\cite{Nomura-K,Kitazawa-H-N} For $\lambda=0$,
with finite $D$ and $\Delta$, the phase diagram has been obtained using
the level-crossing method with twisted boundary
conditions.\cite{Nomura-K,Kitazawa} For example, at $\Delta=1$, a phase
transition from Haldane to large-$D$ phases takes place at $D_{\rm
c}=0.968\pm 0.001$.\cite{Chen-H-S2000,Chen-H-S2003,Chen-H-S2008}

\begin{figure}[t]
\begin{center}
\includegraphics[width=6.0cm]{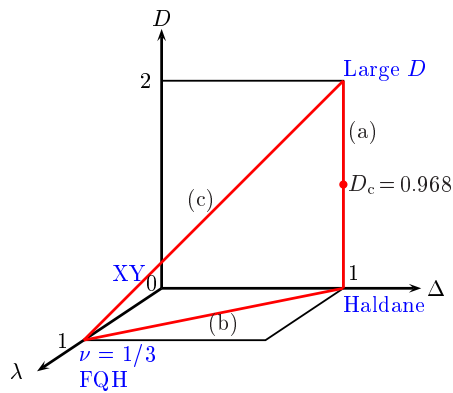}
\end{center}
\caption{(Color online) Parameter space of the model (\ref{model}) connecting three
phases (XY, Haldane, large-$D$ phases) in the $S=1$ quantum spin chain
and the $\nu=1/3$ fractional quantum Hall state.}
\label{fig:threedim}
\end{figure}

In order to analyze the parameter regions beyond the known ones, we 
study the excitation spectra of the system under antiperiodic boundary conditions
using the exact diagonalization.\cite{Nomura-K,Kitazawa} The
antiperiodic boundary conditions have the role of
{\it making the non-degenerate ground states artificially
two-fold degenerate}. In such analysis,
there are four essential excitations that can be used to 
identify four possible phases. By probing the differences
\begin{equation}
\Delta E_i\equiv E_i-E_0,\ (i=1,2,3,4) 
\end{equation}
where $E_0$ is the ground state energy with periodic boundary
conditions, the ground state of the infinite-size system can be
identified according to the lowest excitation in finite-size
systems. This means that phase transition points are
given by level crossings of the two lowest energy levels 
under the twisted boundary conditions.

Discrete symmetry plays an important role in relating the twisted levels
to four physical phases (XY, Haldane, large-$D$, and dimer phases).
According to the valence-bond-solid pictures \cite{AKLT} and
periodicity, the three gapped states under twisted boundary conditions
are classified by space inversion ($\mathcal{P}: S^{\alpha}_{i}\to
S^{\alpha}_{L+1-i}$), spin reversal ($\mathcal{T}: S^{\alpha}_{i}\to
-S^{\alpha}_{i}$),\cite{Kitazawa-N} and translational ($\e^{\mathrm{i}
k}: S^{\alpha}_{i}\to S^{\alpha}_{i+1}$)\cite{wave_number} symmetries as
summarized in Table~\ref{tbl:symmetries}.  In the present system, the
spin-reversal symmetry is always synchronized with the space-inversion
symmetry, hence we refer to them as parity.  The important role of
twisted boundary conditions should be noted here; under periodic
boundary conditions {\it the three gapped states have the same
parity}. In this method, finite-size effects are extremely small even in
small size clusters, since the positions of the level-crossing points
are free from logarithmic corrections.  For example, in case of the
$S=1$ XXZ model ($D=\lambda=0$), there is a level-crossing point between
$\Delta E_1$ and $\Delta E_4$ at $\Delta=0$, which corresponds to the
BKT-type transition between the XY and the Haldane phases (see
Fig.~\ref{dE_XXZ}).\cite{Nomura-K}

\begin{figure}[h]
\begin{center}
\includegraphics[width=7.0cm]{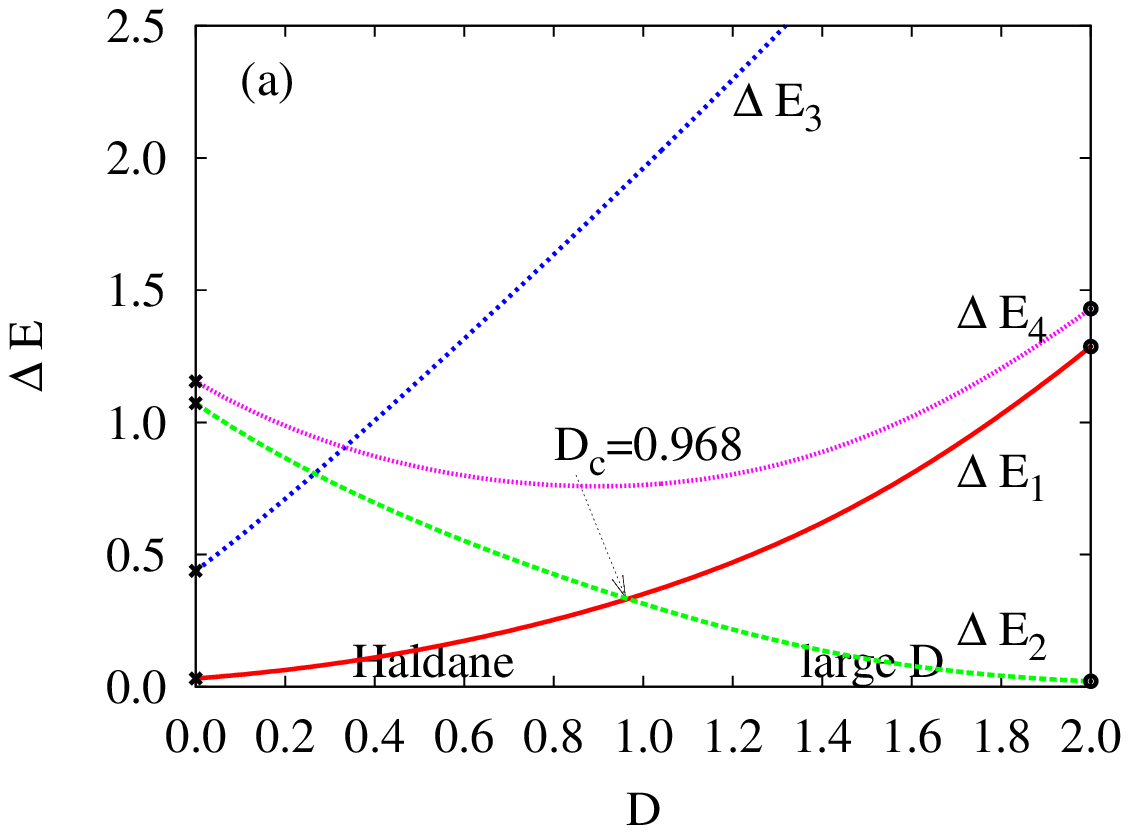}\\
\includegraphics[width=7.0cm]{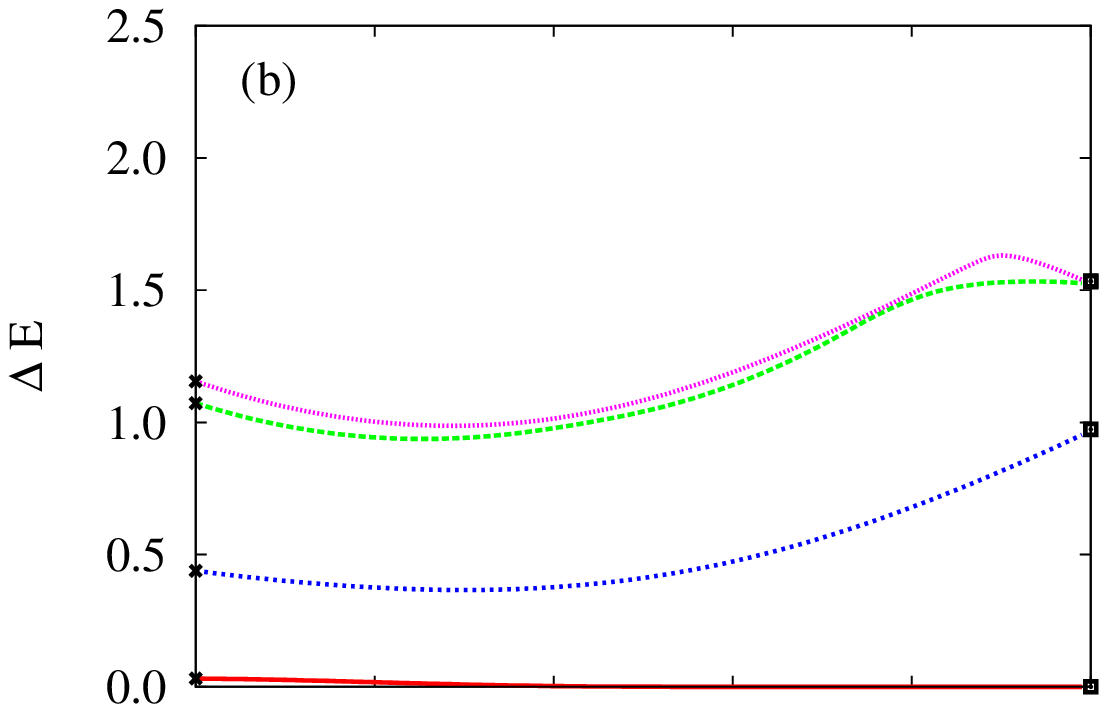}\\
\includegraphics[width=7.0cm]{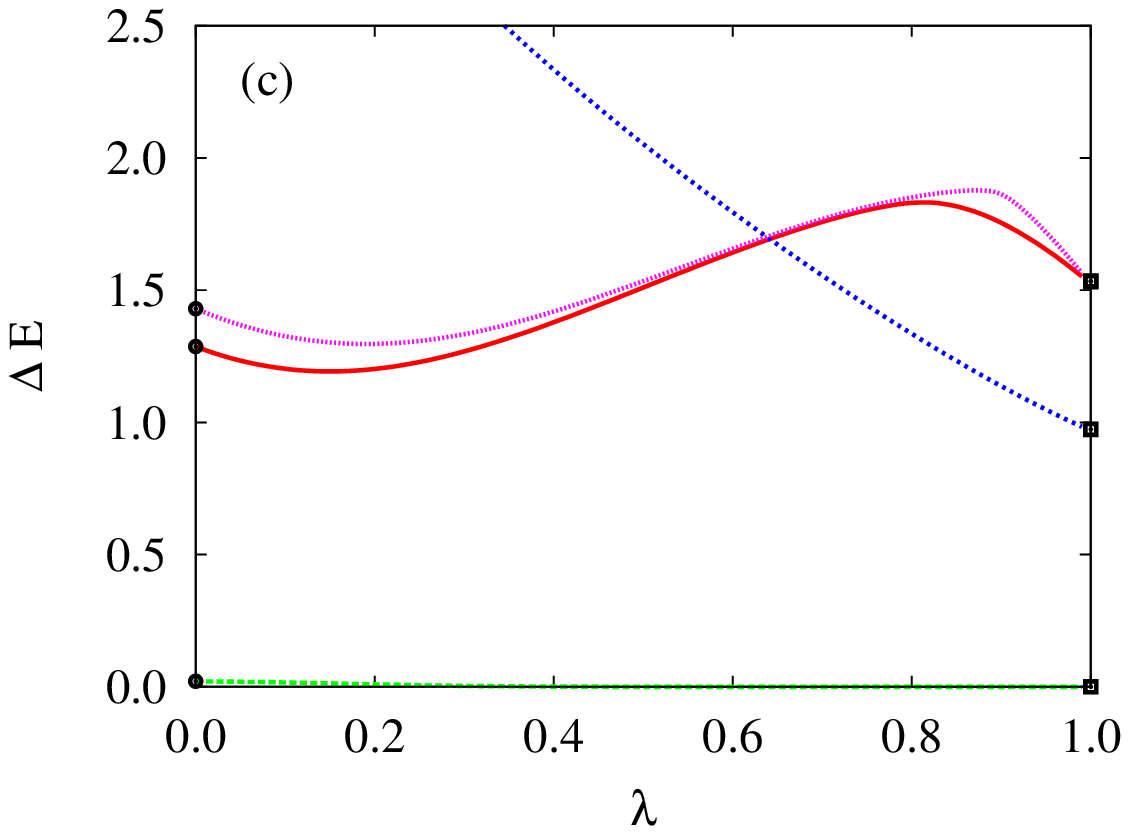}
\end{center}
\caption{(Color online) Excitation spectra of the $N=16$ system under
antiperiodic boundary conditions along the paths (a), (b), and (c) of
Fig.~\ref{fig:threedim}.  In path (a), a level crossing between
different parity ($\Delta E_1$ and $\Delta E_2$) corresponds to the
phase transition point between the Haldane and large-$D$ phases at
$D_{\rm c}=0.968\pm 0.001$.\cite{Chen-H-S2008} In paths (b) and (c),
there are no gap closing points, since there are no level-crossing
points between the lowest two excitations.}\label{fig:dE_ABC}
\end{figure}

According to the conventional classification, the gapped state at the
FQH point ($\Delta=D=0, \lambda=1$) would be expected to belong
either to the Haldane or large-$D$ phases. Therefore, we consider the behavior
of the excitation spectra along the following three paths in the parameter
space (see Fig.~\ref{fig:threedim}): (a)$\Delta=1,\lambda=0$,
(b)$\Delta=1-\lambda,D=0$, and (c)$\Delta=1-\lambda,D=2(1-\lambda)$.
According to the numerical data of the excitation energies obtained by
the exact diagonalization of $N=16$ clusters, there is a phase
transition between Haldane and large-$D$ phases in path (a).  On the
other hand, there is no level-crossing point between the lowest two
spectra in the path (b) and (c), so that the FQH point is adiabatically
connected form both Haldane and large-$D$ phases (see
Fig.~\ref{fig:dE_ABC}).

\begin{figure}[b]
\begin{center}
\includegraphics[width=7.0cm]{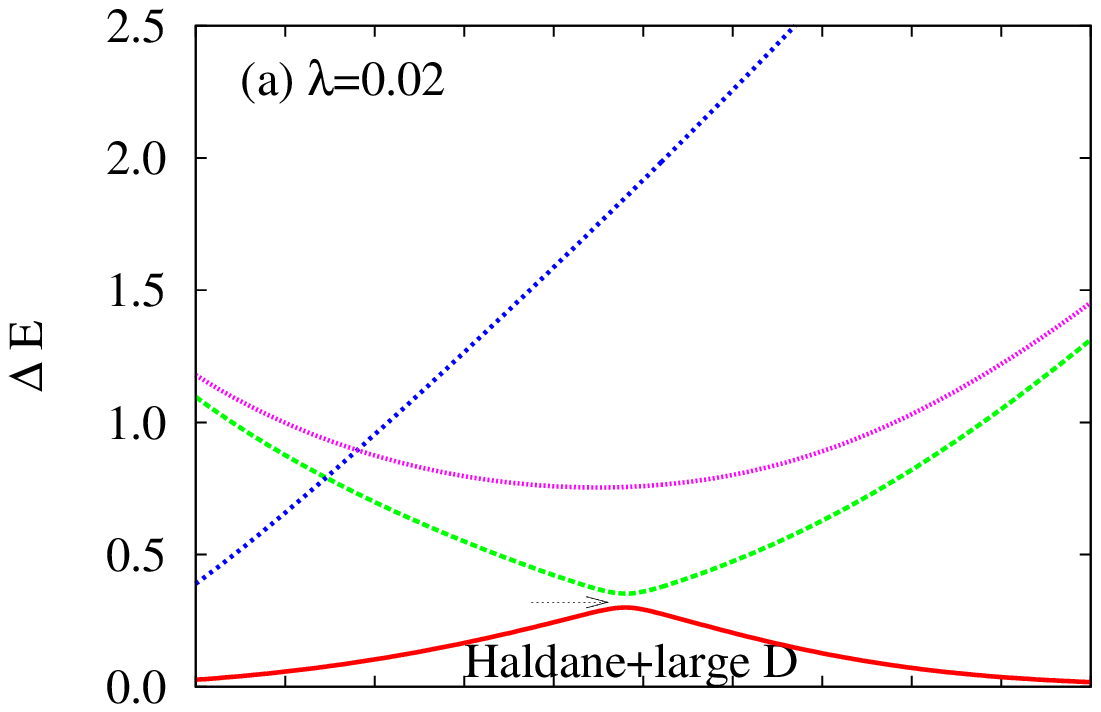}\\
\includegraphics[width=7.0cm]{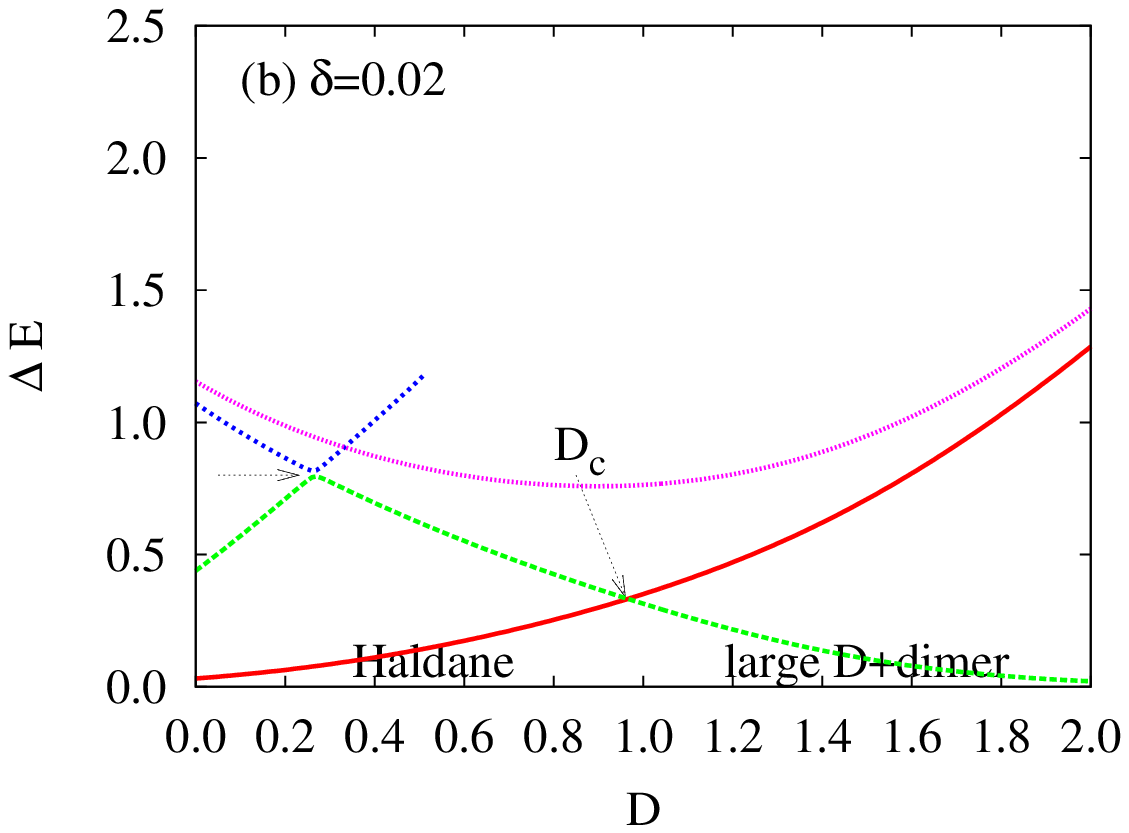}
\end{center}
\caption{(Color online) Excitation spectra with system size $N=16$ around the path (a)
of Fig.~\ref{fig:threedim} with (a) finite $\lambda=0.02$ and (b)
$\delta=0.02$ (bond alternation). In (a), a level repulsion appears
between two spectra $\Delta E_1$ and $\Delta E_2$ due to the breaking
of parity symmetry, while in (b) $\Delta E_1$ and $\Delta E_3$
hybridize due to the breaking of translational symmetry.}\label{dE_Bld}
\end{figure}

The absence of phase transitions can be understood in terms of the
discrete symmetry of the system.  In the excitation spectra along the
path (a) with finite $\lambda>0$, the level crossing between $\Delta
E_1$ and $\Delta E_2$ is absent as shown in Fig.~\ref{dE_Bld}(a).  This
is because there are finite matrix elements between two parity sectors
that were independent in parity-invariant case, and these two energy
levels hybridize. Therefore absence of the level-crossing is due to the
parity symmetry breaking.  Thus the FQH state ($\Delta=D=0$,
$\lambda=1$) belongs to both Haldane and large-$D$ phases.  This
situation is quite similar to the absence of phase transition between
dimer and large-$D$ phases in the $S=1$ bond-alternating Heisenberg
chain with finite dimerization $\delta$.\cite{Chen-H-S2000,Chen-H-S2008}
\begin{equation}
h_{ij}=[1+\delta(-1)^i]\bm{S}_i\cdot\bm{S}_j
 +\frac{D}{2}\left((S_i^z)^2+(S_j^z)^2\right).
\end{equation}
In this case, a similar level repulsion takes place between $E_2$ and
$E_3$ due to the breaking of the translational symmetry (see
Fig.~\ref{dE_Bld}(b)). Arguments for the stability of the Haldane gap
state in terms of symmetry are also discussed in
Refs.~\onlinecite{Berg-T-G-A,Gu-W,Pollmann}.

\subsection{Energy gap}

\begin{figure}[t]
\begin{center}
\includegraphics[width=7.0cm]{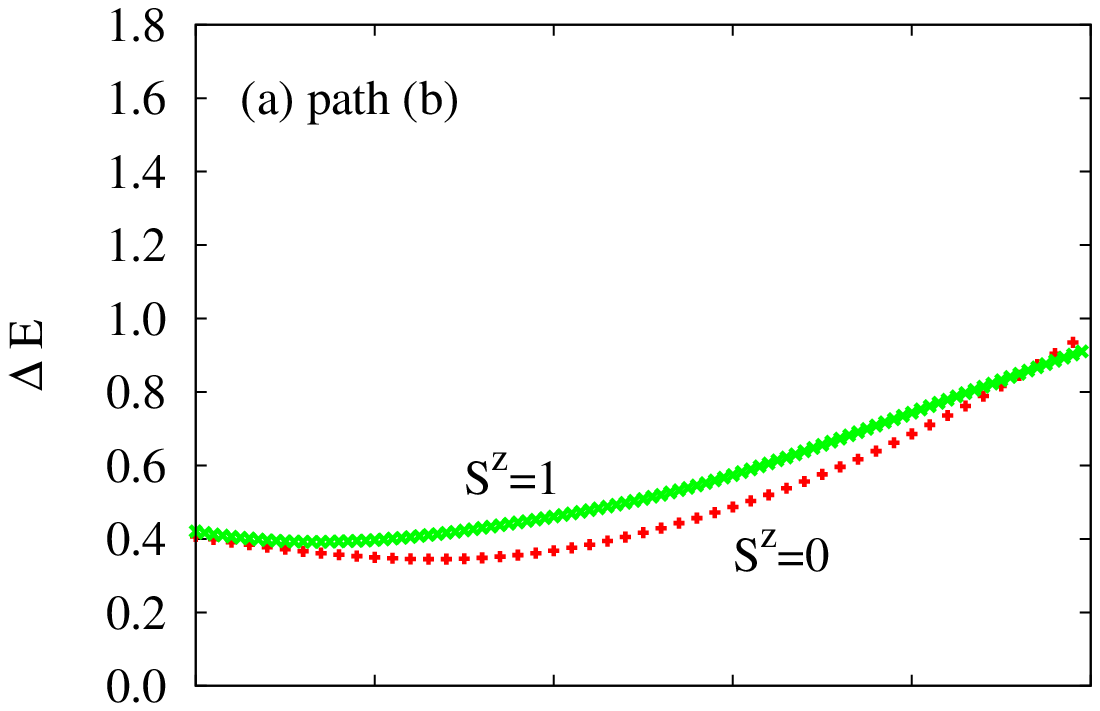}\\
\includegraphics[width=7.0cm]{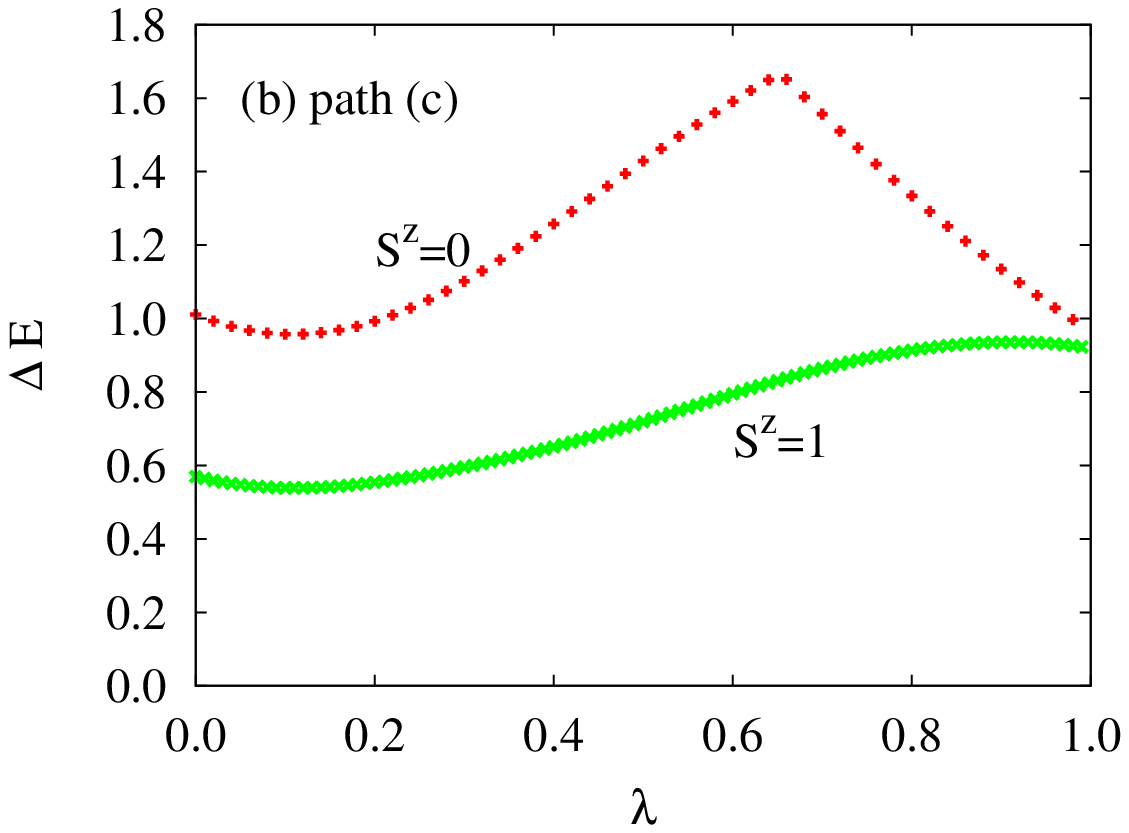}
\end{center}
\caption{(Color online) Extrapolated energy gap for $S^z=0$ and $1$
along the paths (b) and (c) of Fig.~\ref{fig:threedim}.  In path (c),
there is a level crossing in excited states. At the FQH state
($\lambda=1$), $S^z=1$ gap is the lowest which is consistent with the
structure of the excited states with Coulomb
interactions.}\label{fig:energy_gap}
\end{figure}

Let us turn our attention to the behavior of the energy gap for $S^z=0$
and $S^z=1$ excitations along paths (b) and (c) of
Fig.~\ref{fig:threedim}. The energy gaps are obtained by the following
extrapolation function $\Delta E_{g}(N)=\Delta E_{g}(\infty)+A/N+B/N^2$
using the data of the system size $N=8,10,12,14,16$. Especially, for the
$S^z=0$ case, extrapolation of difference between the lowest two
excitation energies under the twisted boundary conditions ($\Delta
E_{1,2,3}$) gives the energy gap with good accuracy.\cite{Nakano-T} We
have checked the validity of our analysis by comparing our result with
the known value of the Haldane gap $E_{g}(\infty)=0.4104\cdots$.  In
Fig.~\ref{fig:energy_gap}, the energy gaps along path (b) and (c) are
shown.  Energy gap for $S^z=2$ which has been omitted is always larger
than that of $S^z=1$.  As we expected, there is no gap closing point along either of the paths. In path (b), the Haldane gap is given by $S^z=0$ gap, and
there is a level-crossing in the excited state close to the FQH point
($\lambda=1$), then the $S^z=1$ state gives the energy gap. In this
sense our $S=1$ model actually give closer description of the $\nu=1/3$
Coulomb state than the the pseudopotential interaction (which has the
Laughlin state as its exact ground state) since the energy gap of the
Coulomb interaction mapped to one dimension has $K=1$ ($S^z=1$) energy
gap, while the pseudopotential interactions has a minimal gap in the
$K=2$ ($S^z= 2$) sector.

\subsection{Order parameters}
In addition to the analysis of energy spectra, we consider the behavior
of string order parameters (SOP's).\cite{Nijs-R} In order to be define
useful SOP's, we extend our $S=1$ model to a $S=1/2$ ladder by making
the following replacements\cite{Kim-F-S-S}:
\begin{equation}
S_{j}^\alpha\to S_{1,j}^\alpha+S_{2,j}^\alpha,
\end{equation}
where $S_{1,j}^\alpha$ and $S_{2,j}^\alpha$ are the $S=1/2$ variables
for the first and the second legs. In this extension, we can introduce
the rung exchange term
\begin{equation}
{\cal H}_{\perp}=\sum_{i=1}^N J_{\perp}\bm{S}_{1,i}\cdot\bm{S}_{2,i},
 \label{rung_exchange_term}
\end{equation}
which at strong ferromagnetic coupling $J_{\perp}\to -\infty$ makes the
ladder system equivalent to the $S=1$ chain. In the present case,
however, the model expressed in the ladder basis commutes with the rung
exchange term of eq.~(\ref{rung_exchange_term}), so that the physics of
the $S=1$ chain is obtained by choosing appropriate value of
$J_{\perp}$.

The purpose of this extension is to introduce the following two
SOP's,\cite{Kim-F-S-S}
\begin{eqnarray}
{\mathcal O}^\alpha_{p}=-\lim_{|k-l|\to\infty}
 \left\langle{\tilde S}_{p,k}^\alpha
  \exp\left[{\rm i}\pi\sum_{j=k+1}^{l-1}{\tilde S}_{p,j}^\alpha\right]
  {\tilde S}_{p,l}^\alpha\right\rangle.
 \label{eqn:SOPs}
\end{eqnarray}
where $\langle\cdots\rangle$ denotes the ground-state expectation value,
and $\alpha=x,y,z$. The composite spin operators for $p= \{{\rm odd},
{\rm even}\}$ are defined by
\begin{eqnarray}
{\tilde S}_{{\rm odd},j}^\alpha=S_{1,j}^\alpha+S_{2,j}^\alpha,\quad
{\tilde S}_{{\rm even},j}^\alpha=S_{1,j}^\alpha+S_{2,j+1}^\alpha.
\label{eqn:composite-spins}
\end{eqnarray}
These two SOP's distinguish between two topologically different
short-range valence bond ground states, namely a state in the
universality class of the Haldane-gapped $S=1$ state described by the
Affleck-Kennedy-Lieb-Tasaki (AKLT) model, and a resonating valence bond
(RVB) state such as the rung dimer state and the large-$D$ phase.  In
terms of these SOP's, the AKLT and the RVB phases are characterized by
${\mathcal O}^\alpha_{\rm odd}\neq0$, ${\mathcal O}^\alpha_{\rm even}=
0$, and by ${\mathcal O}^\alpha_{\rm odd}=0$, ${\mathcal O}^\alpha_{\rm
even} \neq 0$, respectively. Note that ${\mathcal O}^\alpha_{\rm even}$
cannot be defined in the original spin-$1$ system.

\begin{figure}[t]
\begin{center}
\includegraphics[width=7.0cm]{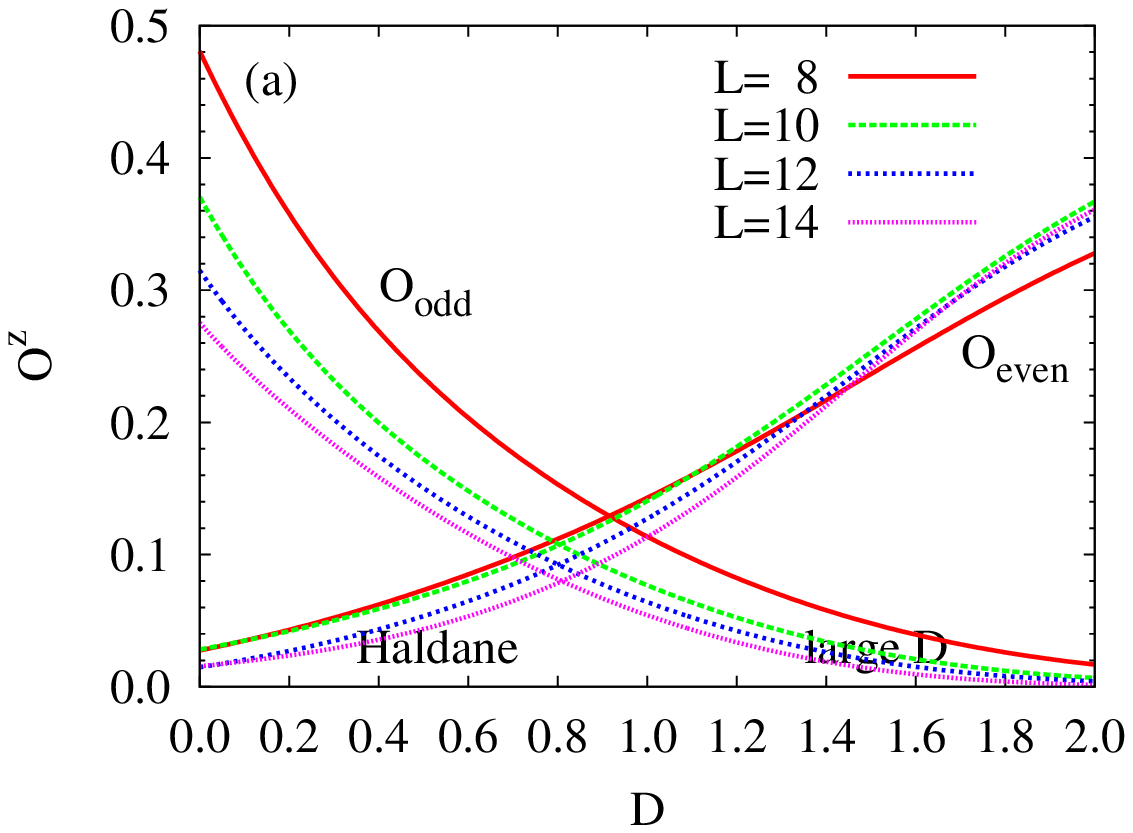}\\
\includegraphics[width=7.0cm]{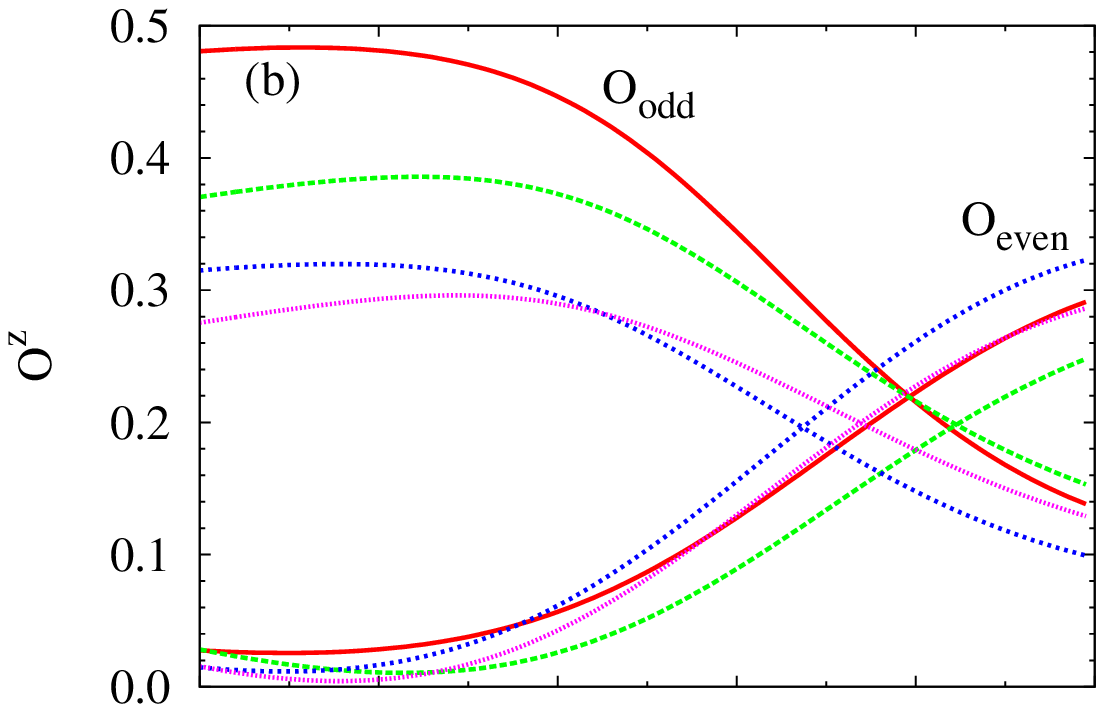}\\
\includegraphics[width=7.0cm]{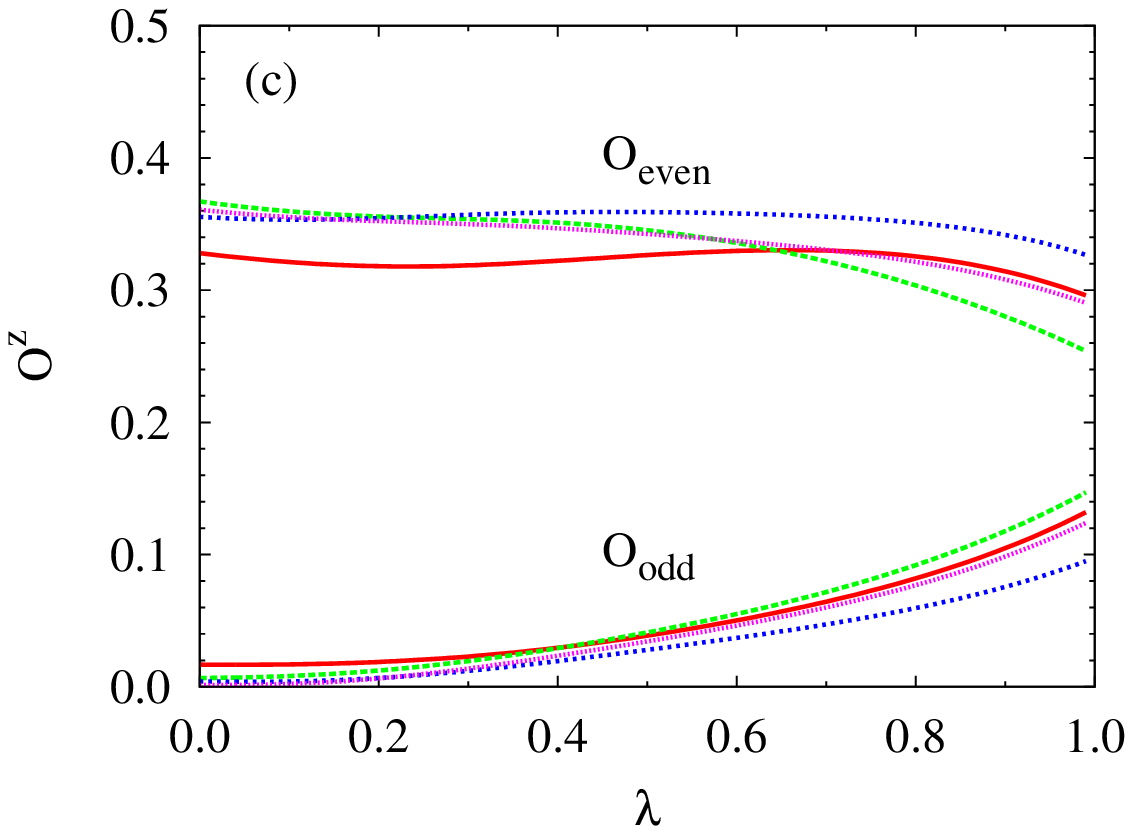}
\end{center}
\caption{(Color online) String order parameters in the ladder system
${\mathcal O}^z_{\rm odd}$ and ${\mathcal O}^z_{\rm even}$ with
$N=10$-$14$, along the paths (a), (b), and (c) (see
Fig.~\ref{fig:threedim}).  The odd-SOP tends to vanish around the
large-$D$ phase and the FQH state while the even-SOP behaves in the
opposite way.}\label{fig:Oz_ABC}
\end{figure}

We calculate the odd and even SOP's in finite-size systems with
$l=k+L/2$, using exact diagonalization, along the three paths (a)-(c)
shown in Fig.~\ref{fig:dE_ABC}. We find that the odd-SOP tends to vanish
for the large-$D$ and the FQH state, while the even-SOP shows the
opposite behavior.  As shown in Fig.~\ref{fig:Oz_ABC} we also find that
crossing points of these two SOP's appear. As discussed in
Ref.~\onlinecite{Nakamura}, in usual spin ladder system with parity
symmetry, a crossing of the SOP's is equivalent to the level-crossing
point of the excitation energies, and gives a transition point between
the AKLT and the RVB phases. However, in the parity-broken system, the
crossing of SOP's does not indicate a phase transition, as discussed
previously. We should also note that the behavior of the even-SOP is
reminiscent of the ``parity order parameter'' of $S=1$ chain discussed
in Ref.~\onlinecite{Berg-T-G-A}, ${\cal O}=
\lim_{|k-l|\to\infty} \langle \exp[{\rm i}\pi\sum_{j=k+1}^{l-1} S_{j}^z]
\rangle$ which has the same bosonized representation as the even-SOP.

The present result indicates that the FQH state around the TT limit is
well characterized by the even-SOP rather than the odd-SOP.  The
large-$D$ phase corresponds to the charge-ordered state in the original
model (\ref{TT_model}), and it well known that the charge-order wave
correlation survives for the system with finite circumference of the
torus $L_1$, so that the even-SOP may remain finite
as $L_1$ is increased. However, the Haldane and the large-$D$ phases
always coexist, and the phase characterized by the even-SOP has the nature
of a ``dual'' Haldane phase, related to the usual Haldane phase by a
duality analogous to the Kramers-Wannier relation in the two-dimensional
Ising model.\cite{Nakamura} Therefore, the present system has
properties characteristic of the Haldane phase, 
provided that the energy gap does not
close. Note that the parity symmetry is absent also 
in the original
one-dimensional model with long range interactions (\ref{TT_model}).

Recently, a classification of topological phases in one and two
dimensions has been suggested which is based on degeneracies in the
entanglement spectrum.\cite{Li-H,Pollmann} This approach may also shed light on
the relationship between quantum Hall systems and quantum spin
systems.

\section{conclusion}\label{sec4}

In conclusion, we have discussed the relationship between the $\nu=1/3$
FQH state in the vicinity of the TT limit and $S=1$ spin chains with
Haldane gap.  In the TT limit, the ground state is charge-ordered and
corresponds to the large-$D$ phase of the $S=1$ spin chain.  Away form
the TT limit, the system retains characteristics of the large-$D$ phase,
but the Haldane phase also coexists due to the broken parity
symmetry. It is plausible that features of the pure Haldane phase may
become more pronounced as the circumference of the torus $L_1$ is
increased beyond the range of applicability of our analysis (see
Fig.~\ref{fig:pdiagram}). This scenario is supported by the observation
in Ref. \onlinecite{Girvin-A} that the off-diagonal long-range order in
the Laughlin state is very similar in nature to the string order in the
Haldane phase. 

As outlined in this work, the present analysis of the $\nu=1/3$ also be
extended to general (even) odd-denominator filling fractions, $\nu$, by mapping
to (half-)integer-$S$ spin chains. It is well known that the both the gapped
Haldane and large-$D$ phases only exist for odd-integer spin chains,
and the present work signals their relevance to the odd-denominator rule
in the hierarchy of fractional quantum Hall states.

\section{Acknowledgments}

We acknowledge many discussions with K.~Hida, A.~Karlhede, A.~L\"auchli,
H.~Nakano, M.~Oshikawa, and S.~Ryu. M.~N. acknowledges the visitors
program at the Max-Planck-Institut f\"{u}r Physik komplexer Systeme,
Dresden, Germany, and support from Global Center of Excellence Program
``Nanoscience and Quantum Physics'' of the Tokyo Institute of Technology
by MEXT.



\begin{thebibliography}{99}
\bibitem{worry}
	One might worry that the three-fold degeneracy of the TT state
	would cause trouble as they would map onto three different spin
	configurations for each of the three groupings into spin
	sites. However, these states are not connected by the
	Hamiltonian as the correspond to different $K$-values, hence it
	does make sense to define one spin subspace corresponding to
	each of the three degenerate ground states (and their respective
	low energy excitations).
\bibitem{wave_number}
	 Large-$D$ and dimer states can be classified by the wave
	 number due to difference of the periodicity.
\end{thebibliography}
\end{document}